\documentclass[
reprint,
superscriptaddress,
amsmath,amssymb,
aps,
prl
]{revtex4-2}

\usepackage[T1]{fontenc}
\usepackage[utf8]{inputenc}
\usepackage{graphicx}
\usepackage{dcolumn}
\usepackage{bm}
\usepackage{hyperref}
\hypersetup{hidelinks}

\begin{document}
\raggedbottom
\title{Moir\'e-Induced Interlayer Electron-Phonon Coupling in Graphene/hBN Superlattices}

\author{Dasol Kim}
\thanks{These authors contributed equally to this work.}
\affiliation{Department of Physics, Massachusetts Institute of Technology, Cambridge, Massachusetts, USA}

\author{Zhengguang Lu}
\thanks{These authors contributed equally to this work.}
\affiliation{Department of Physics, Massachusetts Institute of Technology, Cambridge, Massachusetts, USA}

\author{Tonghang Han}
\thanks{These authors contributed equally to this work.}
\affiliation{Department of Physics, Massachusetts Institute of Technology, Cambridge, Massachusetts, USA}

\author{Guorui Chen}
\thanks{These authors contributed equally to this work.}
\affiliation{Materials Science Division, Lawrence Berkeley National Laboratory, Berkeley, CA, USA}
\affiliation{Department of Physics, University of California at Berkeley, Berkeley, CA, USA}

\author{Tianyi Han}
\affiliation{Department of Physics, Massachusetts Institute of Technology, Cambridge, Massachusetts, USA}

\author{Jixiang Yang}
\affiliation{Department of Physics, Massachusetts Institute of Technology, Cambridge, Massachusetts, USA}

\author{Bing Cheng}
\affiliation{Department of Physics, Massachusetts Institute of Technology, Cambridge, Massachusetts, USA}

\author{Zhenxian Liu}
\affiliation{Department of Physics, University of Illinois Chicago, Chicago, IL, USA}

\author{Christopher Homes}
\affiliation{National Synchrotron Light Source II, Brookhaven National Laboratory, Upton, NY, USA}

\author{G. L. Carr}
\affiliation{National Synchrotron Light Source II, Brookhaven National Laboratory, Upton, NY, USA}

\author{Jeong Min Park}
\affiliation{Department of Physics, Massachusetts Institute of Technology, Cambridge, Massachusetts, USA}

\author{Qihang Zhang}
\affiliation{Department of Physics, Massachusetts Institute of Technology, Cambridge, Massachusetts, USA}

\author{Lili Jiang}
\affiliation{Department of Physics, University of California at Berkeley, Berkeley, CA, USA}

\author{Hongyuan Li}
\affiliation{Materials Science Division, Lawrence Berkeley National Laboratory, Berkeley, CA, USA}

\author{Kenji Watanabe}
\affiliation{Research Center for Electronic and Optical Materials, National Institute for Materials Science, 1-1 Namiki, Tsukuba 305-0044, Japan}

\author{Takashi Taniguchi}
\affiliation{Research Center for Materials Nanoarchitectonics, National Institute for Materials Science, 1-1 Namiki, Tsukuba 305-0044, Japan}

\author{Pablo Jarillo-Herrero}
\affiliation{Department of Physics, Massachusetts Institute of Technology, Cambridge, Massachusetts, USA}

\author{Feng Wang}
\affiliation{Materials Science Division, Lawrence Berkeley National Laboratory, Berkeley, CA, USA}
\affiliation{Department of Physics, University of California at Berkeley, Berkeley, CA, USA}
\affiliation{Kavli Energy NanoSciences Institute at the University of California, Berkeley, CA, USA}

\author{Long Ju}
\email{longju@mit.edu}
\affiliation{Department of Physics, Massachusetts Institute of Technology, Cambridge, Massachusetts, USA}

\begin{abstract}
Electron-phonon coupling (EPC) plays a vital role in the electrical and optical properties of materials, ranging from superconductivity to charge dynamics. In moir\'e superlattices, out-of-plane vibrations can modulate the moir\'e potential and enhance interlayer EPC. Here we report the first direct observation of a resonance arising from coupling between moir\'e electronic excitations and phonons in graphene-based moir\'e superlattices. Using Fourier-transform infrared photocurrent spectroscopy, we detect the nominally silent hexagonal boron nitride (hBN) out-of-plane optical (ZO) phonon as a pronounced absorption peak. The phonon interferes with moir\'e flat-band excitations to form a characteristic Fano resonance. Displacement field-dependent measurements confirm that the feature originates from coupling to the moir\'e electronic continuum rather than simple zone folding. We further observe similar resonances in devices with different graphene layer numbers and twist configurations. This enhanced EPC provides a pathway for understanding and controlling emerging quantum phenomena, such as superconductivity, and engineering novel optoelectronic devices.
\end{abstract}

\maketitle
Stacking two-dimensional (2D) materials into moir\'e superlattices provides a powerful way to engineer novel electrical \cite{yankowitz_emergence_2012,dean_hofstadters_2013,ponomarenko_cloning_2013,hunt_massive_2013} and optical \cite{wu_topological_2017,jin_observation_2019,tran_evidence_2019,alexeev_resonantly_2019,seyler_signatures_2019} phenomena of quantum materials. In particular, the 2D moir\'e potential leads to flat electronic minibands, which provides a new platform to study correlated electron phenomena featuring novel superconductivity \cite{cao_unconventional_2018,chen_signatures_2019,yankowitz_tuning_2019,lu_superconductors_2019,park_tunable_2021,hao_electric_2021,xia_superconductivity_2025,guo_superconductivity_2025}, correlated insulating states \cite{chen_evidence_2019,cao_correlated_2018,chen_electrically_2021,wang_correlated_2020,cao_tunable_2020,liu_tunable_2020,tang_simulation_2020,xu_correlated_2020,regan_mott_2020}, and topological states \cite{serlin_intrinsic_2020,sharpe_emergent_2019,chen_tunable_2020,xie_fractional_2021,park_observation_2023,lu_fractional_2024}. The moir\'e potential depends sensitively on the spatial overlap between electronic wavefunctions in adjacent layers for a given atomic configuration. Therefore, an out-of-plane lattice vibration in one layer can strongly couple to electrons in a neighboring layer by dynamically modulating the interlayer separation and thus the moir\'e potential. This interlayer EPC channel is typically weak in non-moir\'e materials but can be amplified in moir\'e systems by the enhanced density of states and the strong sensitivity of miniband wavefunctions to small potential modulations.

Previously, interlayer EPC has been investigated in various van der Waals heterostructures, including studies using Raman spectroscopy in WSe$_2$/hBN heterostructures \cite{jin_interlayer_2017} and large-twist-angle bilayer graphene \cite{eliel_intralayer_2018,righi_resonance_2013,campos-delgado_raman_2013}. However, these systems lack the long-period moir\'e potential necessary to reconstruct the electronic spectrum into flat minibands. Consequently, interlayer EPC enabled or strongly enhanced by moir\'e effects remains largely unexplored. Recent studies in TMD moir\'e systems have shown that moir\'e reconstruction can strongly reshape phonon spectra and phonon-related optical responses \cite{rahman_extraordinary_2022,quan_phonon_2021}. In parallel, photoemission \cite{chen_strong_2024} and quantum twisting spectroscopies \cite{birkbeck_quantum_2025} in graphene-based moir\'e systems have revealed strong coupling between moir\'e electronic states and phonon channels, providing emerging access to EPC in moir\'e materials. Together, these advances show that moir\'e band reconstruction can substantially reshape EPC and motivate an optical spectroscopic study for EPC, potentially involving a resonance where a flat-band electronic continuum couples to a discrete lattice vibration.

\begin{figure*}[t]
\centering
\includegraphics[width=0.9\textwidth]{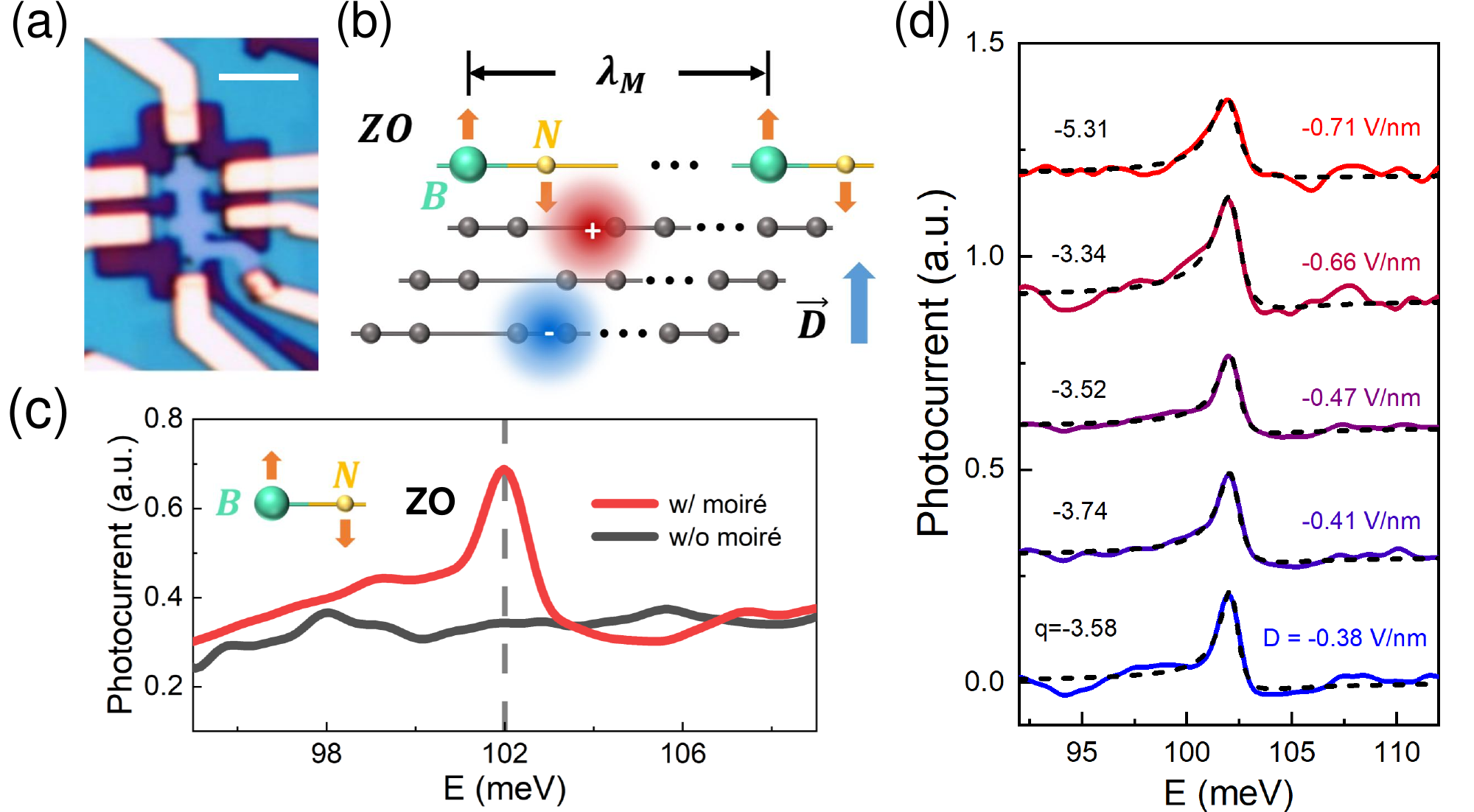}
\caption{Interlayer coupling between hBN ZO phonon and EHPs in TLG. (a) Optical micrograph of an ABC TLG/hBN device. Scale bar: 3 $\mu$m. (b) Illustration of the interlayer EPC in TLG/hBN with a moir\'e wavelength $\lambda_M$. The ZO phonon in hBN carries zero in-plane electrical dipole and remains silent in both Raman and infrared absorption processes. Through dynamic modulation of the moir\'e potential, the hBN ZO phonon hybridizes with electron (blue cloud) and hole (red cloud) pairs in neighboring graphene layers and gains electrical dipole. (c) Comparison of photocurrent spectra measured in TLG devices. A pronounced peak near 102 meV appears only in the moir\'e device, while it is strongly suppressed in the device without moir\'e, indicating that the otherwise silent hBN ZO phonon becomes active through moir\'e-enabled coupling to electron-hole pairs in TLG. (d) Photocurrent spectra at different gate displacement field $D$. The asymmetric line shape is due to interference of the renormalized ZO phonon with EHPs at the immediate neighborhood, which can be well described by a Fano line shape with a $q$ factor represented by the dashed curves. The smooth background is removed for the Fano fitting. Spectra are shifted along the y-axis for clarity.}
\label{fig:1}
\end{figure*}

Here we report the observation of a resonance arising from coupling between moir\'e electronic excitations and hBN phonons in graphene-based moir\'e superlattice materials. Different from photoluminescence and Raman scattering, which detect phonons through inelastic scattering processes involving higher-energy electronic transitions \cite{jin_interlayer_2017,eliel_intralayer_2018,righi_resonance_2013,rahman_extraordinary_2022}, we used FTIR photocurrent spectroscopy \cite{ju_tunable_2017} to probe the absorption of infrared photons by the hybridized electron-phonon states at the phonon energy (see Supplemental Material \cite{SM}\nocite{zhang_bridging_2019,miller_observing_2009,zhang_direct_2009,slawinska_energy_2010,shi_gate-dependent_2014,jiang_infrared_2007,ribeiro-palau_twistable_2018,klots_probing_2014}). In a perfectly aligned ABC trilayer graphene (TLG)/hBN moir\'e device with $\lambda_M \approx$ 14.5 nm \cite{chen_signatures_2019} (Fig. 1(a)), we observed a Fano resonance between the dressed hBN ZO phonon and resonant electron-hole pairs (EHPs), with its broadening and energy tunable by displacement field $D$. We then explored the dependence of EPC on $\lambda_M$, and experimentally demonstrated $\lambda_M$-dependent interlayer EPC in MLG/hBN. In an hBN-encapsulated 1.8$^{\circ}$ tBLG device, we observed a stronger hBN ZO phonon feature than that in TLG/hBN and MLG/hBN. We model the interlayer EPC in the TLG/hBN moiré system with calculable wavefunctions and obtain results consistent with the experimental observations, demonstrating quantitatively the role of the moiré superlattice.

\begin{figure*}[t]
\centering
\includegraphics[width=0.9\textwidth]{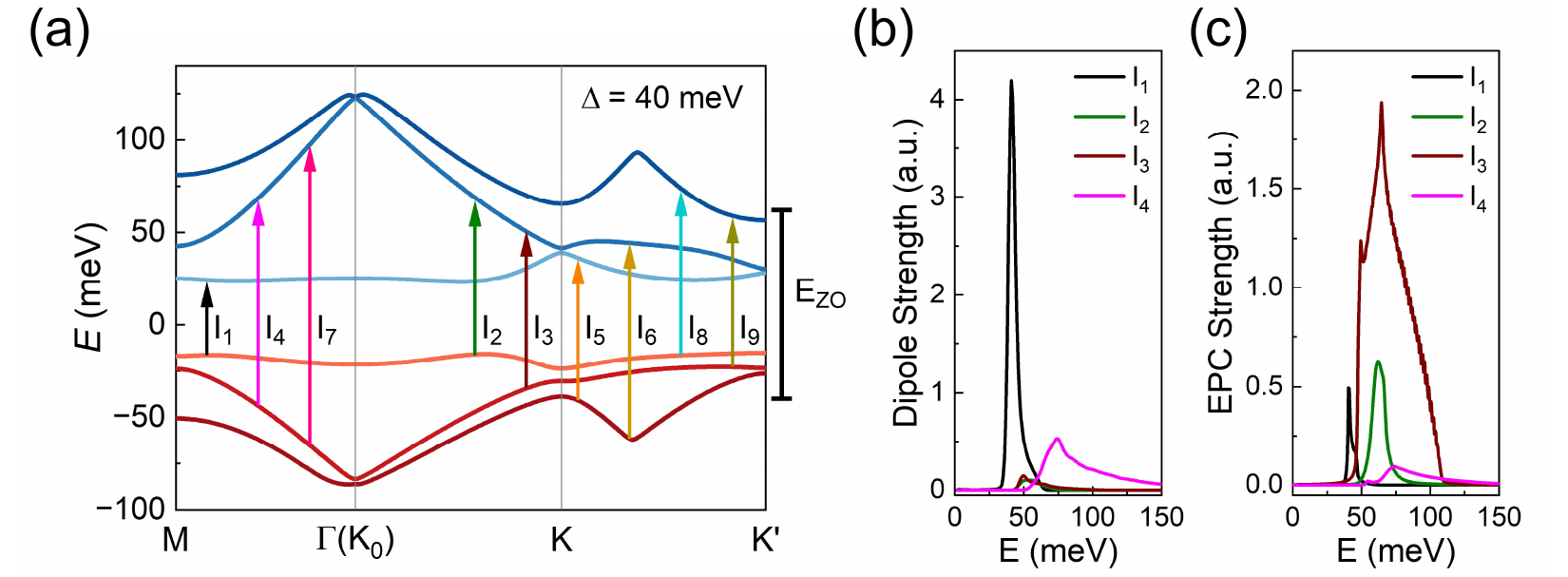}
\caption{The theoretical model to explain interlayer EPC in TLG/hBN. (a) Band structure of TLG in the mini-Brillouin zone under a moir\'e potential at interlayer potential difference $\Delta$ = 40 meV. $I_1$, $I_4$, $I_7$ label EHPs that can be optically excited in the absence of moir\'e potential, while $I_2$, $I_3$, $I_5$, $I_6$, $I_8$ and $I_9$ label EHPs that are optically active only in the presence of moir\'e potential. (b) Calculated optical dipole (oscillator) strength for $I_1$-$I_4$ as a function of excitation energy at $\Delta$ = 40 meV under a static moir\'e potential. (c) Corresponding calculated EPC strength for $I_1$-$I_4$. $I_1$ and $I_4$ show large oscillator strength but weak EPC, whereas $I_2$ and $I_3$ exhibit the opposite trend.}
\label{fig:2}
\end{figure*}

The ZO phonon in hBN is illustrated in Fig. 1(b). Due to its out-of-plane nature, it is usually silent in both infrared absorption and Raman spectra under in-plane polarized illumination \cite{geick_normal_1966,serrano_vibrational_2007}. In graphene/hBN moir\'e superlattices, however, the ZO mode can gain an in-plane electric dipole through hybridization with optically active electron-hole pairs in graphene. As a result, three key observations were made in our experiment. Firstly, the ZO phonon appears as a peak in the photocurrent spectrum of TLG/hBN moir\'e device under in-plane polarized light illumination, as shown in Fig. 1(c). In contrast, no peak at this energy is observed in the moir\'eless TLG device. Secondly, the EPC-induced spectral change has an asymmetric line shape (Fig. 1(d)), well described by the Fano line shape \cite{fano_effects_1961,tang_tunable_2010}
\begin{equation}
\Delta A = A_{e}\left(\frac{\left[q \cdot \gamma + \left(E-E_{\Omega}\right)\right]^{2}}{\gamma^{2}+\left(E-E_{\Omega}\right)^{2}} - 1\right)
\end{equation}
where $E_{\Omega}$ and $\gamma$ are the energy and broadening of the renormalized ZO phonon, respectively; $q$ is the Fano asymmetry parameter and $A_e$ sets the resonance amplitude. A smooth background is removed in Fig. 1(d) before fitting (see Supplemental Material \cite{SM}). The asymmetric Fano resonance peak implies quantum interference between the dressed ZO phonon and EHPs near $E \approx 102\,$meV. Thirdly, the ZO phonon feature is tunable with gate displacement field $D$. For $D$ < 0, the phonon peak shifts slightly while its broadening increases gradually with $|D|$. Data for $D$ > 0 are included in the Supplemental Material \cite{SM}.

The observed interlayer EPC can be understood intuitively: As the ZO phonon is excited, the out-of-plane displacements of B and N atoms modulate the wavefunction overlap and the amplitude of moir\'e potential with the B and N contributions adding up constructively (see Supplemental Material \cite{SM}), thereby renormalizing the electron wavefunctions in TLG. In return, optically active
EHPs that are sensitive to the moiré potential hybridize with the hBN ZO phonon and dress it with an in-plane electrical dipole.

To go beyond phenomenological Fano fitting, we model the interlayer EPC in three steps. (1) We compute the low-energy band structure and electron wavefunctions of TLG/hBN in the presence of a static moir\'e potential $V_{M} = \sum_{j = 1}^{6}{V_{j} \cdot e^{iG_{j} \cdot r}}$ using a tight-binding model \cite{chen_evidence_2019,chittari_gate-tunable_2019,wallbank_generic_2013,jung_ab_2014,abergel_infrared_2013}. (2) From the eigenstates, we further calculate the EPC matrix element $M = \left\langle e|H_{el - phonon}|h \right\rangle $ and the EPC strength $V_{E} = \ \sum_{\alpha,k}^{}\delta_{E_{\alpha,k},E}{{|M}_{\alpha,k}|}^{2}$. (3) Using the EPC matrix elements, we calculate the renormalized phonon energy and broadening. Details are provided in the Supplemental Material \cite{SM}.

Figure 2(a) shows the calculated low energy band structure at interlayer potential difference $\Delta$ = 40 meV. The moir\'e potential folds electron and hole states into the mini-Brillouin zone and forms multiple moir\'e mini-bands. Compared to materials without moir\'e, this unique feature of moir\'e superlattices results in many EHPs that can hybridize with the ZO phonon at momentum $k$ = 0. This is the only phonon relevant to the optical absorption process discussed here. The nine lowest energy EHPs $I_1$-$I_9$ as labeled in Fig. 2(a) can be grouped into two groups: Group 1 ($I_1$, $I_4$, and $I_7$) connect electron and hole states that are at the same momenta in the absence of moir\'e, while Group 2 ($I_2$, $I_3$, $I_5$, $I_6$, $I_8$, and $I_9$) connect those at different momenta in the absence of moir\'e. The two groups of EHPs have distinct properties, as shown in Fig. 2(b) and 2(c). For clarity, we show $I_1$-$I_4$ as representative examples. Group 1 transitions exhibit much stronger optical oscillator strength (Fig. 2(b)), whereas Group 2 transitions, which are unique to moir\'e materials, dominate the interlayer EPC strength (Fig. 2(c)).

The contrasting behavior of Group 1 and Group 2 EHPs can be seen by comparing $I_1$ and $I_2$. When $V_{M} = 0$, $I_1$ connects initial state $| h \rangle$ and final state $| e \rangle$ at the same crystal momentum, while $I_2$ connects states whose momenta differ by a moir\'e reciprocal vector $G_{j}$. Since the electron-photon interaction Hamiltonian carries negligible momentum, the electrical dipole $\sim$ $\left\langle e|H_{el - photon}|h \right\rangle$  is finite for $I_1$ but zero for $I_2$. When $V_{M}$ is turned on, $I_2$ gains a small dipole through hybridization of states separated by $G_{j}$. Consequently, the electrical dipole and oscillator strength of $I_1$ remains much larger than those of $I_2$. In contrast, the EPC matrix element $M = \left\langle e|H_{el - phonon}|h \right\rangle$ preferentially couples electronic states $| e \rangle$ and $| h \rangle$ whose momenta differ by $G_{j}$ since the electron-phonon interaction Hamiltonian $H_{el - phonon}$ is proportional to the moir\'e potential modulation $V_{M} = \ \sum_{j = 1}^{6}{V_{j} \cdot e^{iG_{j} \cdot r}}$. Accordingly, the Group 2 transitions---which are nominally forbidden in the absence of a superlattice---become the dominant channels for enhancing the interlayer EPC strength.

\begin{figure}
\centering
\includegraphics[width=\columnwidth]{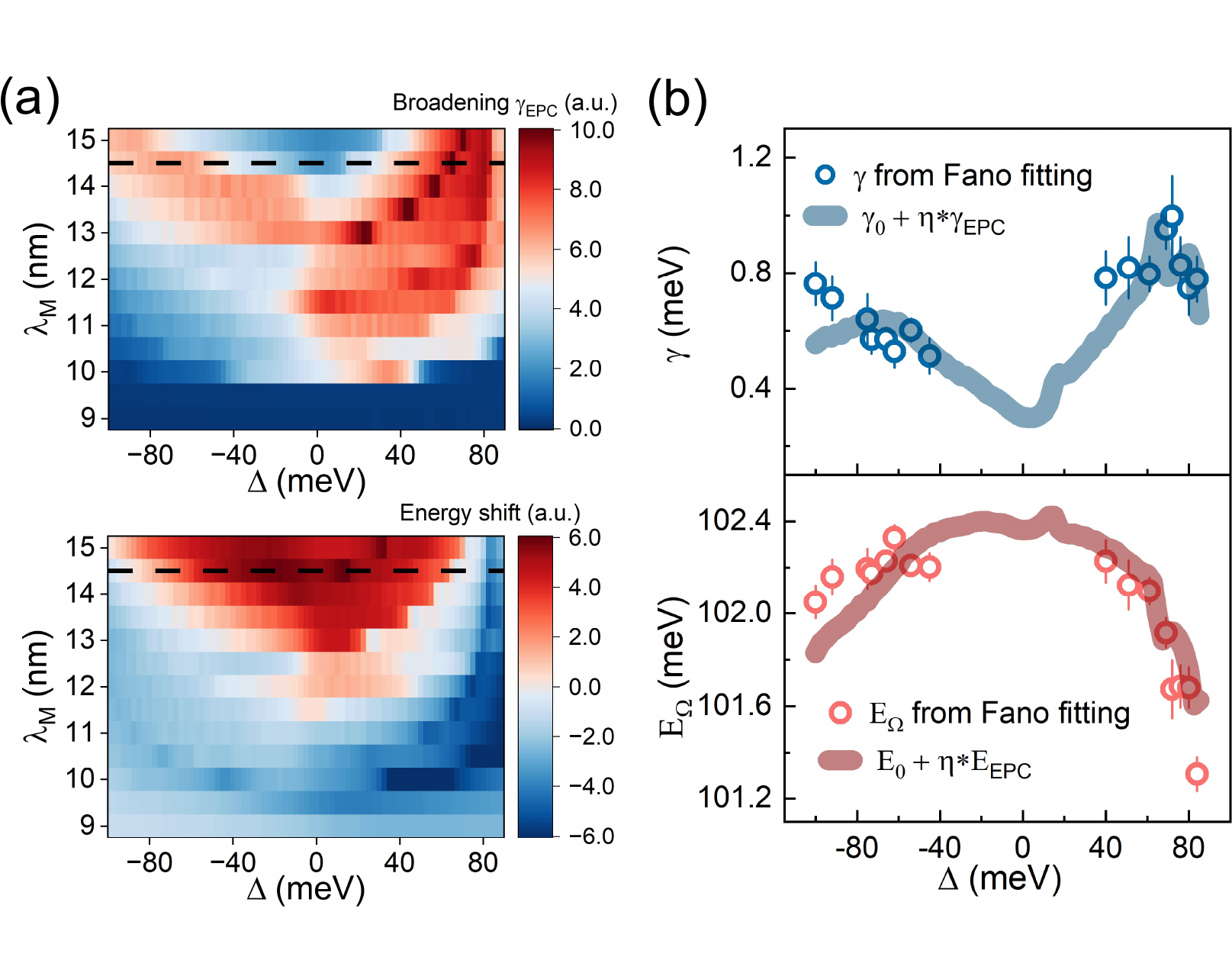}
\vspace{-2.5\baselineskip}
\caption{Calculated and experimental broadening and energy shift of renormalized hBN ZO phonon in TLG/hBN. (a) The calculated broadening $\gamma_{EPC}$ and energy shift $E_{EPC}$ of hBN ZO phonon through coupling with EHP states in TLG. Both quantities vary as a function of $\Delta$ and $\lambda_M$. The dashed line corresponds to $\lambda_M$ = 14.5 nm in our device. (b) Comparisons
between fitted peak broadening and energy of renormalized
hBN ZO phonon and theoretical calculations. The curves
correspond to line-cuts at the dashed lines in (a), offset by the
intrinsic broadening $\gamma_0$ and energy $E_0$ of the hBN ZO phonon.
}
\label{fig:3}
\vspace{-1.0\baselineskip}   
\end{figure}

Using the calculated EPC matrix elements for $I_1$-$I_9$, we calculate the EPC-induced broadening and energy shift of the hBN ZO phonon to compare with the fitting results. Figure 3(a) shows the results as a function of both $\Delta$ and $\lambda_M$. There is one unknown parameter $\eta$ in the EPC calculation, which quantifies the gradient of the moir\'e potential (see Supplemental Material \cite{SM}). Using $\eta$ as an adjustable parameter, with an intrinsic broadening $\gamma_0$ = 0.12 meV and an intrinsic (bare) ZO-phonon energy $E_0$ = 101.93 meV, consistent with previously reported values of $\sim$102 meV \cite {jin_interlayer_2017,jung_vibrational_2015,hu_far-field_2016,li_phonon_2023}, we find a good agreement between the fitting results and calculated values as shown in Fig. 3(b). The correspondence between $D$ and $\Delta$ is described in the Supplemental Material \cite{SM}. Because $D$ tunes the electronic spectrum without changing the lattice periodicity, this agreement indicates that the phonon feature originates from coupling to the moir\'e electronic continuum rather than from zone folding. We note that interlayer EPC is typically much weaker than intralayer EPC due to the reduced spatial overlap between electronic states and lattice vibrations across the interface. The EPC-induced broadening observed here is of the same order of magnitude as that reported for intralayer EPC in graphene-based systems \cite{tang_tunable_2010,kuzmenko_gate_2009,li_structure-dependent_2012}, reflecting a substantial enhancement of interlayer EPC by the moir\'e superlattice.

An intuitive picture further clarifies the peak-like phonon response and its asymmetric line shape. As shown in Fig. 2(b) and Fig. S4 in the Supplemental Material \cite{SM}, EHPs $I_1$--$I_9$ occupy a wide energy range of more than 100 meV and can all couple to the hBN ZO phonon. EHPs close to the phonon energy contribute most strongly to the phonon broadening and Fano interference, whereas EHPs away from resonance contribute to the optical dipole of the dressed phonon. Importantly, the off-resonant contributions from electronic transitions below and above the phonon energy have opposite signs. Over the displacement-field range in our experiment, the dominant electronic resonances lie predominantly below the ZO-phonon energy, so their contributions tend to add rather than cancel, giving the dressed phonon a substantial optical dipole. This explains why the hBN ZO phonon remains predominantly peak-like over a broad $D$ range. In contrast, when a strong electronic resonance is centered near the phonon energy, contributions from the two sides can partially cancel, reducing the phonon optical dipole and producing a dip-like Fano response ($ q \approx 0$) \cite{fano_effects_1961,kuzmenko_gate_2009,li_structure-dependent_2012}. Near the renormalized phonon energy, resonant EHPs interfere with the dressed phonon and produce the asymmetric Fano line shape \cite{fano_effects_1961,kuzmenko_gate_2009,li_structure-dependent_2012}.

In addition to the gate-controlled $\Delta$, the EPC can also be tuned by $\lambda_M$, as shown by our calculation in Fig. 3(a). In TLG/hBN the interlayer EPC is expected to be negligible at small $\lambda_M$: as $\lambda_M$ decreases, relevant EHP energies move away from the fixed ZO phonon energy due to the enlarged mini-Brillouin zone, reducing EPC-induced broadening and energy shift. For the same reason, in a previously studied hBN-encapsulated AB-stacked bilayer graphene \cite{ju_tunable_2017}, where no transport signature of a moir\'e superlattice was observed, the hBN ZO phonon feature was not observed. Due to the difficulty in fabricating TLG/hBN devices, we instead test the $\lambda_M$ dependence in MLG/hBN devices and observe a systematic weakening of the phonon feature with decreasing $\lambda_M$ (see Supplemental Material \cite{SM}). This trend supports additional tuning of interlayer EPC by $\lambda_M$ in 2D moir\'e superlattice materials.

\begin{figure*}[t]
\centering
\includegraphics[width=0.9\textwidth]{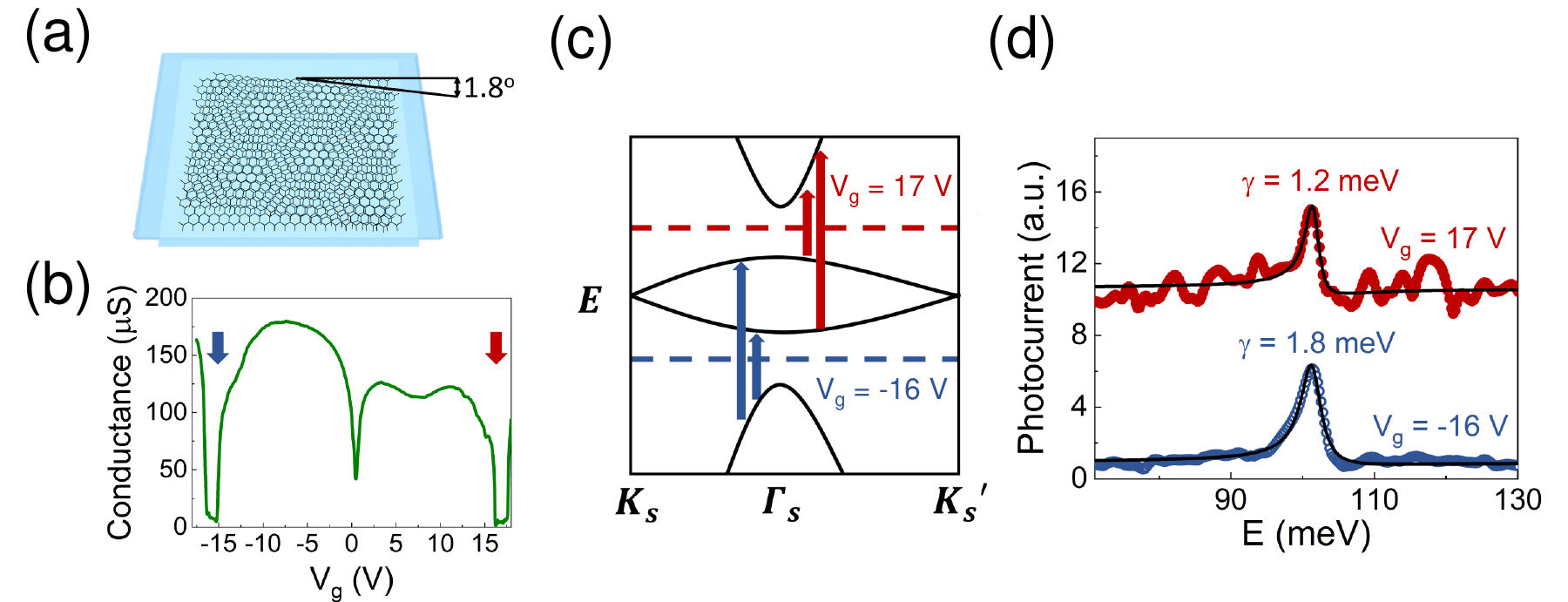}
\caption{Interlayer coupling between hBN ZO phonon and EHPs in tBLG. (a) Illustration of tBLG encapsulated by hBN flakes. The twist angle between two layers of graphene is 1.8$^{\circ}$, while the twist angle between graphene and hBN is unknown. (b) Gate-dependent conductance of tBLG device where blue and red arrows indicate when the Fermi level is in moir\'e gaps. (c) Illustration of the tBLG band structure where EHPs indicated by arrows can couple to the hBN ZO phonon. (d) Photocurrent spectra of this device. The spectrum at $V_g$ = 17 V is shifted up along the y-axis by 9 units for clarity. When the Fermi level is in either moir\'e gap, the photocurrent spectrum is dominated by the hBN ZO phonon feature. We extracted $\gamma$ = 1.8 meV at $V_g$ = -16 V and 1.2 meV at $V_g$ = 17 V through Fano fitting.}
\label{fig:4}
\end{figure*}

Given the flexibility in assembling multilayer van der Waals heterostructures, different moir\'e superlattices could coexist in the same synthetic material. To examine whether the interlayer EPC also occurs in other graphene-based moiré systems, we measured an hBN-encapsulated tBLG device using the same photocurrent spectroscopy measurement. There is a 1.8$^{\circ}$ twist angle between the two layers of graphene as illustrated by Fig. 4(a). We note that the alignment between graphene and the top/bottom hBN flake was not purposely controlled during fabrication. In the gate-dependent device conductance shown in Fig. 4(b), we observed two dips that are symmetric about the charge-neutrality dip at $V_g \approx$ 0 V. They correspond to when the Fermi level is tuned into the moir\'e gaps \cite{cao_superlattice-induced_2016}. EHPs excited across these two moir\'e gaps are illustrated by arrows in Fig. 4(c). Fig. 4(d) shows the photocurrent spectra of this device which clearly show the hBN ZO phonon feature. From Fano fitting, we extract $\gamma \sim$ 1.8 meV, which is 2-3 times bigger than that in TLG/hBN and MLG/hBN, indicating stronger interlayer EPC in this tBLG/hBN device. We do not yet have a complete microscopic picture of interlayer EPC in hBN-encapsulated tBLG, where the twist angle between the graphene and hBN layers is large. One possibility is that the hBN ZO vibration breaks the symmetry between the top and bottom graphene layers and modulates the tBLG minibands. A quantitative description would require both graphene/graphene and graphene/hBN moir\'e potentials, together with lattice relaxation \cite{nam_lattice_2017,woods_commensurateincommensurate_2014,yoo_atomic_2019}.

In summary, our observation of a moir\'e-induced interlayer EPC in several graphene/hBN moir\'e materials reveals a dynamic coupling between electrons in graphene and phonons in neighboring hBN layers. While the microscopic mechanism is established here for ABC-TLG/hBN through detailed modeling, the observations in multiple graphene/hBN moiré systems suggest that this coupling is broadly present.
These observations point to two immediate directions for further explorations. Firstly, EPC is key to charge transport process such as temperature-dependent conductivity and formation of Cooper pairs in conventional and unconventional \cite{lee_interfacial_2014} superconductivity. Although our experiment here focuses on a particular
phonon mode in hBN, it is intriguing to explore interlayer
EPC between electrons in graphene and other phonon modes to understand their roles in superconductivity and linear-in-$T$ resistance in graphene-based moiré superlattice materials \cite{cao_unconventional_2018,chen_signatures_2019,cao_strange_2020,polshyn_large_2019}. Secondly, strong interlayer EPC may also enable phononic devices such as phonon lasers in BLG or ABC-TLG. Compared with the previously proposed BLG scheme based on a $\sim$200 meV G-mode phonon \cite{tang_tunable_2010}, the much lower energy of the hBN ZO phonon makes it easier to achieve strong coupling in a realistic gate range. The additional
phonon broadening due to interlayer EPC suggests an efficient phonon-emission channel under optical pumping. Together, these results point to the possibility of driving a nonequilibrium ZO-phonon population by optically
pumping the electron-hole pairs in graphene/hBN moiré superlattices.

\bigskip
\noindent\textit{Acknowledgments}---We acknowledge discussions with L. Levitov, R. Ashoori, M. Koshino, and J. Xu. We thank Y. H. Zhang for providing the codes for band structure calculation. This study was supported by the U.S. Department of Energy, Office of Science, Basic Energy Sciences, Materials Sciences and Engineering Division under Grant No. DE-SC0025325. Device fabrication was performed at the Harvard Center for Nanoscale Systems and MIT.nano. T.H.H. acknowledges support from a MathWorks Fellowship. G.C. and F.W. acknowledge support from the U.S. Department of Energy, Office of Science, Basic Energy Sciences, Materials Sciences and Engineering Division under Contract No. DE-AC02-05-CH11231 within the van der Waals heterostructure program KCWF16. Work in the P.J.-H. group was primarily supported by the Army Research Office MURI W911NF2120147, the MIT Microsystems Technology Laboratories Samsung Semiconductor Research Fund, the National Science Foundation (DMR-1809802), the Gordon and Betty Moore Foundation EPiQS Initiative through Grant No. GBMF9463, and the Ramon Areces Foundation. This research used resources of the 22-IR-1 (FIS) and 22-IR-2 (MET) beamlines of the National Synchrotron Light Source II, a U.S. Department of Energy (DOE) Office of Science User Facility operated for the DOE Office of Science by Brookhaven National Laboratory under Contract No. DE-SC0012704. Operations at FIS are funded by NSF Synchrotron Earth and Environmental Science (SEES, EAR-2223273) and DOE/NNSA Chicago/DOE Alliance Center (CDAC, DE-NA0004153) cooperative agreements.

\bigskip
\noindent\textit{Author contributions}---L.J. conceived and supervised the experiment. D.K., Z.G.L., T.Y.H., J.Y., Q.Z., B.C., Z.X.L., C.H., and G.L.C. performed the photocurrent spectroscopy measurements. T.Y.H. and G.C. fabricated the samples with help from L.L.J., H.L., and J.M.P. K.W. and T.T. grew hBN single crystals. T.H.H., T.Y.H., D.K., and L.J. performed the theoretical analysis. T.H.H., D.K., and L.J. wrote the paper with input from all authors.

\bibliographystyle{apsrev4-2}
\bibliography{epc_final_Aug13}
\end{document}